# Chromatin assortativity: Integrating epigenomic data and 3D genomic structure

**Vera Pancaldi[1,*], Enrique Carrillo-de-Santa-Pau[1], Biola Maria Javierre[2], David Juan[1], Peter Fraser[2], Mikhail Spivakov[2], Alfonso Valencia[1,3] and Daniel Rico[1,3,*]**

[1]Structural Biology and BioComputing Programme, Spanish National Cancer Research Centre (CNIO), Madrid, Spain

[2]Nuclear Dynamics Programme, The Babraham Institute, Cambridge, United Kingdom

[3]Co-senior authors

*Corresponding authors: vpancaldi@cnio.es; drico@cnio.es

# Abstract

**Background**

The field of 3D chromatin interaction mapping is changing our point of view on the genome, paving the way for new insights into its organization. Network analysis is a natural and powerful way of modelling chromatin interactions. Assortativity is a network property that has allowed social scientists to identify factors that affect how people establish social ties. We propose a new approach, using chromatin assortativity to integrate the epigenomic landscape of a specific cell type with its chromatin interaction network and thus investigate which proteins or chromatin marks mediate genomic contacts in the nucleus.

**Results**

We use high-resolution Promoter Capture Hi-C and Hi-Cap data as well as ChIA-PET data from mouse embryonic stem cells to investigate promoter-centered chromatin interaction networks. We then calculate the presence of a collection of 78 epigenomic features in the chromatin fragments constituting the nodes of the network. We use the chromatin assortativity approach to estimate the association of these epigenomic features to the topology of four different chromatin interaction networks and use high assortativity to identify features localized in specific and connected areas of the network. We found Polycomb Group proteins and associated histone marks as the features with the highest chromatin assortativity in promoter-centred networks. We then ask which features distinguish contacts amongst promoters from contacts between promoters and other genomic elements. Remarkably, we observe higher chromatin assortativity of the actively elongating form of RNA Polymerase 2 (RNAPII) compared to inactive forms only in interactions between promoters and other elements. This suggests a prominent role for active elongation in distal regulatory interactions.

**Conclusions**

Contacts amongst promoters and between promoters and other elements have different characteristic epigenomic features. Using chromatin assortativity we identify a possible role for the elongating form of RNAPII in mediating interactions among promoters, enhancers and transcribed gene bodies. Our approach facilitates the study of multiple genome-wide epigenomic profiles, considering network topology and allowing for the comparison of any number of chromatin interaction networks.

# Background

Advances in chromatin interaction mapping have allowed us to refine our vision of the genome, leading us to a more realistic well organized tension globule picture with extrusions of chromatin loops [1, 2]. The resolution of available contact maps has increased from a megabase to less than a kilobase in just 5 years [3–10]. However, our understanding of what determines three-dimensional structure and of its functional importance remains limited. Starting from the first papers modelling DNA as a polymer and the genome as a polymer globule [1, 2, 11], scientists have been looking for a connection between the chromatin contact configuration and the regulation of gene expression [12–14]. It is now accepted that gene regulation happens as much through distal enhancer elements as through proximal promoters and the distinction between promoters and enhancers has itself been put to test [15, 16].

The combination of chromatin capture experiments with next-generation sequencing (NGS) has enabled the characterization of chromatin contacts at an unprecedented level of detail. Different techniques yield different views of the genome. High-throughput conformation capture (HiC) is an unbiased approach that allows to investigate the three-dimensional structure of the genome of given cell types [3, 9], even in single cells [17] during differentiation processes [10, 18–20] and across species [21, 22]. The HiC technique assays, in principle, all versus all chromosomal contacts, requiring very high sequencing coverage and making it very costly and practically almost impossible to achieve saturating coverage. Alternative approaches allow exploration of the contacts of a subset of genomic regions, with higher resolution at the same cost. For example, Chromatin Interaction Analysis by Paired-End Tag sequencing (ChIA-PET) [23] analyzes only those interactions that are mediated by a protein of interest by pulling down only the interacting fragments that include this protein.

Recently, other capture approaches were developed that enable to selectively enrich for genome wide interactions involving, at least on one end, specific regions of interest. For example, Capture HiC was recently used to identify the chromatin interactions involving colorectal cancer risk loci [24]. A similar approach is used in promoter capture HiC (PCHi-C) [8], which detects both promoter-promoter interactions and interactions of promoters with any other non-promoter regions. These interactions are therefore identified irrespective of target promoter activity and across the whole range of linear genomic distances between fragments. HiCap [7] is a similar approach to detect promoter-centered chromatin interactions. The two methods provide a complementary view of chromatin interactions as PCHi-C yields larger fragments (average fragment size 5Kb) and longer interaction ranges (on average 250Kb), whereas HiCap has better resolution (average fragment size < 1Kb) but less coverage of long range interactions. Thanks to these new techniques, we can now use interactions between non-coding parts of the genome and genes to interpret the wealth of disease associated genomic variation data which were so far unexplained [24–26].

The increasing availability of 3D interaction datasets for multiple cell types and organisms has prompted the development of multiple data processing approaches. There are important factors that need to be taken into account in these analyses. One is the detection of biologically significant interactions from the background noise of interactions purely due to linear proximity of the two fragments on the genome; another is the averaging effect that is produced by the heterogeneity of contacts in different cells [27]. While various methods for normalizing and detecting signals in HiC related datasets have been developed [28–30], downstream interpretation of the resulting contact maps represents a significant problem. Moreover, to this day, no single unified standards are available for these types of data, hindering the direct comparison between the chromatin structure in different cell types, species or conditions [28]. However, the field is moving fast, as shown by the recent focus on unravelling the 4D nucleome, that is the internal organization of the nucleus in space and time, even at the resolution of single cells [31, 32].

Given the complexity of these datasets, it is intuitive and useful to represent them as networks, in which each chromatin fragment is a node and each edge (link) represents a significant interaction between two chromatin fragments. This framework allows us to study the properties of the 3D chromatin structure using tools from network theory. The booming field of network science provides a useful toolbox and different metrics that can be used to compare and interpret chromatin contact networks from a more global point of view. For example, one can identify the most connected nodes or look for functional relationships between nodes that interact more than expected by chance [33].

A few previous papers have dealt with network analysis approaches applied to chromatin interaction networks [33–37], with the aim of unravelling general principles of 3D chromatin organization. For example, In the pioneering work by Sandhu et al. [35], the chromatin interaction network is constructed starting from RNAPII ChIA-PET performed in mouse embryonic stem cells (mESCs) to obtain a single large connected component. An accurate network analysis revealed the functional organization of different chromatin communities. A similar analysis, performed on the budding yeast chromatin interaction network, showed that cohesin mediates highly interconnected interchromosomal subnetworks (cliques) which are stable and have similar replication timing [33].

In this work, we aim to establish which properties of, or factors bound to the DNA can be associated to specific types of 3D chromatin contacts. To this end, we project the linear chromatin context information directly onto the 3D network, preserving its topology. We focus our analysis on mESCs as chromatin interactions for this cell type have been assayed by multiple techniques and a very comprehensive epigenetic characterization is available. We study interaction networks derived by state-of-the-art PCHi-C in mESCs, in which we quantify the assortativity of 78 chromatin features (3 cytosine modifications, 13 histone modifications and 62 chromatin-related proteins binding peaks [38]).

In social sciences, assortativity is used to measure the extent to which similar people tend to connect with similar people [39, 40]. Whereas in society it is easy to imagine which principles might lead people from the same ethnic origin or cultural background to establish social ties, we are still investigating principles that organize chromatin in the nucleus. We borrow the concept of assortativity making an analogy between social networks and chromatin contact networks and introduce the concept of chromatin assortativity (ChAs). This global measure identifies to what extent a property of a chromatin fragment is shared by fragments that interact preferentially with it. If a feature appears to be localized in specific well-connected areas of the network, it will be characterised by having high ChAs. Identifying features with high ChAs can thus lead us to candidates for factors that might mediate chromatin contacts. This would be an important step forward in elucidating the organizing principles inside the nucleus and furthering our understanding of the mechanistic basis of genome regulation.

Polycomb Group (PcG) proteins and associated marks have the highest ChAs values, imposing themselves as the factors that are more strongly related with chromatin structure, as recently suggested [5, 20, 41]. Through this novel analysis, we also gain insight regarding different RNA Polymerase II (RNAPII) variants as important players shaping the 3D chromatin structure. More specifically, we note a different configuration of actively elongating RNAPII forms in promoter-other end contacts compared to non-elongating RNAPII variants. This finding is confirmed in three independent datasets and it suggests that actively elongating RNAPII is involved in the contact between regulatory elements and their targets.

# Results

## The Chromatin Interaction Network

To assemble the chromatin interaction network, we used the recent PCHi-C dataset in mESCs from Schoenfelder et al. [8], including interactions amongst promoters and between promoters and other genomic elements. The PCHi-C data was processed using the CHiCAGO algorithm. CHiCAGO is a HiC data processing method that filters out contacts that are expected by chance given the linear proximity of the interacting fragments on the genome and takes into account the biases introduced by the capture step used in the PCHi-C approach [29]. The network based on the significant interactions detected by CHiCAGO has 55,845 nodes and 69,987 connections (see Methos and **Figure S1, Additional File 1**). Of these interactions, 20,523 interactions connect a promoter fragment with another promoter fragment (P-P edges) and 49,464 interactions connect promoters with non-promoter "other end" fragments (P-O edges).

As in many networks, we can observe a main large connected component (LCC) that consists of 35,293 nodes (63% of total nodes) joined by 52,984 edges (76% of total edges) (**Figure S1**). There are 264 disconnected components with more than 10 nodes and about 4,000 additional small components. Each chromatin fragment has an average of 2.5 neighbours with each promoter interacting with three non-promoter elements on average.

## Epigenomic features associated to chromatin fragments participating in 3D contacts

For each fragment in the PCHi-C network, we mapped a large set of 78 epigenomic features [38]. These features included cytosine modifications, histone marks and ChIP-seq peaks of chromatin-related proteins, such as transcription factors and members of chromatin complexes including cohesin, CTCF, PcG and different RNAPII variants (**Additional File 2**). For each chromatin fragment we calculate the fraction covered by peaks of a specific feature and we define the abundance of each feature as the average of this value over all fragments in the network (see Methods). **Figure 1A** shows the fraction of fragments covered by EZH2 binding sites. We noticed the strong accumulation of the nodes that have binding sites for this PcG factor in specific regions of the network. Strikingly, this co-localization of the signal is observed despite the low overall prevalence of EZH2 binding in the fragments (only 10% of fragments have some overlap with EZH2 peaks, Table S2). **Figure 1B** shows the HoxA cluster region on chromosome 6. In this region, we observe that fragments connected by long-range interactions tend to have similar values of EZH2, with EZH2 peaks having similar heights on pairs of connected fragments. We therefore set out to investigate and quantify the extent to which connected nodes in the whole network have similar values for EZH2 and the other 77 epigenomic features. High similarity of values in interacting nodes could suggest a role for some features in mediating these contacts.

**Definition of Chromatin Assortativity (ChAs)**

We propose an approach to identify epigenomic features that can be associated to 3D chromatin contacts. This involves measuring the extent to which neighbouring network nodes have similar epigenomic features, using chromatin assortativity (ChAs). Assortativity, also called homophily, is the propensity for interacting nodes to have similar values [40] (see Methods). ChAs is defined as the assortativity of abundance levels of one specific epigenomic feature on the chromatin interaction network.  In practical terms, it is the correlation of abundance of a single feature measured across all pairs of neighbours in the network. As a correlation coefficient, ChAs values range between -1 and 1. ChAs can therefore be used to identify features that are found in fragments that are globally connected in the network or to distinguish different type of fragments that tend to interact with each other. To aid the interpretation of these values, we can consider the three scenarios depicted in a schematic scatter plot of ChAs versus abundance (**Figure 1C**):

1) Fragments that have a certain value of the epigenomic feature (that is a certain proportion of the fragment is covered by peaks of that feature) predominantly interact with other fragments which have similar values of the same feature, but not with other fragments. In this case the ChAs for that feature will be positive (ChAs>0). This situation would indicate that this feature is potentially associated to chromatin contacts.

2) Alternatively, there might be no relationship between the values of the feature on fragments and values on their neighbours. In this case we will have ChAs=0. This can happen either when the feature values do not have anything to do with the contacts, or when the feature values are very homogeneous in the network: either the feature is low on all fragments (as would be the case for a very rare chromatin mark, or high on all fragments (as would be the case for ubiquitous chromatin marks). This produces low variability of abundance across nodes and hence the correlation of these values in neighbouring nodes measured by ChAs tends to be 0.

3) Finally it could be that fragments that have high values of a given feature frequently interact with fragments with low values of that same feature. In this case we will have a negative ChAs (ChAs<0). This suggests that a set of genomic regions with the feature tend to interact in the network mostly with fragments of a different kind.

For this reason, it is important to consider the abundance of a feature (defined above as the fraction of fragment covered by the feature averaged over fragments) together with the ChAs value. In our EZH2 example, the abundance of this feature is 0.027 and its ChAs value is 0.34, which demonstrates how a fairly rare feature can be assortative.

To summarize, firstly we are interested in features that have high positive ChAs, as this signifies that the mark appears to be localized in specific connected areas of the network. These features are thus very probably involved in the chromatin contacts. Secondly, we are looking for features with negative ChAs, which should be typical of one subclass of fragments that frequently interact with a different subclass of fragments. In this case, ChAs can be used to detect features that distinguish multiple chromatin fragment types.

A recent cohesin ChIA-PET dataset [42] allows us to illustrate the characteristics and biological interpretation of ChAs. Dowen et al. reported interactions with pulldowns of the SMC1 cohesin unit in mESCs [42]. We therefore proceeded to measure abundance and ChAs of SMC1 in this dataset, obtaining a fairly high value of abundance (0.27, mean of all features 0.09) and a low value of ChAs (0.09, mean o all features 0.28). This is expected due to the strong enrichment of fragments for presence of this protein (98% of fragments have an SMC1 peak). This enrichment makes all fragments have similar proportions covered by the SMC1 feature, hence driving down the ChAs value. CTCF, on the contrary, shows an almost 3-fold increase in ChAs (0.29 vs. 0.09 of SMC1) and only a 1.2% increase in abundance (0.33 vs 0.27 of SMC1) as compared to SMC1. These results, suggest that the subset of cohesin-bound fragments that have in addition CTCF bound tend to interact preferentially with each other. In summary, using this well understood dataset we showed that ChAs is a measure that combines the presence of peaks in different interacting fragments and the topology of the chromatin interaction network. ChAs can thus detect differences and biases in the different types of chromatin interaction networks and identify the chromatin features playing important roles in 3D structure in the cases where these are not known *a priori*.

## ChAs of chromatin features in the mESC chromatin interaction network detected by PCHi-C

We calculated ChAs for the 78 chromatin features in the entire PCHi-C network and compared these values with the corresponding abundance (**Figure 2A**). The PcG proteins (EZH2, PHF19, RING1B, SUZ12, CBX7) and histone marks associated with them (H3K27me3, H2Aub1) have the highest ChAs values (ranging from 0.2 to 0.35, mean of all features 0.08, **Figure 2A**), suggesting that this complex might be involved in establishing the 3D structure of chromatin in mESC cells. This confirms and extends results observed for the Hox gene clusters [8, 20, 41]. RNA polymerase II (RNAPII) also has high ChAs, especially the variant implicated in transcriptional elongation (ChAs of RNAPII-S2P=0.23, **Figure 2A**). Two features with high abundance that also have high ChAs are H3K4me1, found on regulatory distal regulatory elements and H3K36me3, marking transcribed gene bodies. On the other hand, H3K4me3, a modification associated with active promoters, is a very abundant mark (fourth most abundant, abundance=0.12, mean of all features 0.02), but it has low ChAs (0.04).

We verified that ChAs is robust to random removal of edges in the network, such that our results do not depend on the completeness and accuracy of the chromatin interaction network (See **Text S1 and Figure S2, Additional File 1**). Moreover, we have ensured the significance of ChAs for at least 72% of the features using a randomization that preserves network topology and overall feature abundance, as well as using an alternative approach preserving the features' spatial distribution (See **Text S1 and Figures S3**, **Additional File 1**). We have also verified that ChAs values are in general not affected by removing short-range contacts that might produce similarity of abundance values in neighbouring fragments (**Figures S4-5, Additional File 1**). Finally, comparison of ChAs with other network measures demonstrates that it is a complementary method to identify important features (See **Text S2-3, Figures S6-7**, **Additional File 1**).

In summary, the ChAs of an epigenomic feature is a useful global measure that relates feature abundance at interacting fragments with the underlying interaction network topology. In the next section, we will compare the ChAs values calculated on different chromatin interaction networks.

## Chromatin assortativity in an additional Promoter Capture HiC and ChIA-PET datasets

To test to what extent chromatin interaction network properties depend on the experimental protocol and signal detection algorithm, we took advantage of an alternative promoter interaction dataset in mESCs. Sahlén et al. applied HiCap (a promoter capture method similar to PCHi-C) also to mESCs, identifying interactions involving promoters [7]. Using contacts amongst promoters and between promoter and non-promoter fragments from the Sahlén et al. dataset yields a network of 87,823 nodes with 173,801 interactions (of which 19,309 promoter nodes and 82,659 P-P interactions). The HiCap technique is complementary to PCHi-C, since a different enzyme is used for the restriction step, generating shorter interaction fragments compared to PCHi-C (median size 599 bp vs 3,953 bp for PCHi-C). The shorter fragments produce a higher resolution picture of contacts between nearby fragments, at the expense of reduced coverage of large-scale interactions. Visualizing the network shows that the largest connected component is comparatively smaller than in PCHi-C, encompassing 9.6% of the total nodes and 12.8% of the total connections (**Figure S8**, **Additional File 1**).

We analysed the HiCap network in combination with the 78 chromatin features previously introduced. We repeated the calculation of ChAs of the chromatin features using the HiCap network as described above for the PCHi-C network. We directly compared the ChAs values for all features between PCHi-C and HiCap networks and found that, overall, they are highly correlated (Pearson's R = 0.67, p-value =2.99 x $10^{-11}$, **Figure 2B and Figure S8D,E, Additional file 1**). For example, the PcG components are confirmed amongst the features with the highest ChAs, as was observed in the PCHi-C analysis, together with RNAPII, especially the S2P variant (**Figure S8E, Additional File1**).

In summary, we have shown that ChAs is a useful metric to detect those epigenomic features that might be more influential in promoter-centered chromatin interaction networks and that the ChAs measurements are rather independent of the underlying experimental protocol. A comparison with a contact map that is not enriched for contacts involving promoters was performed using the previously-mentioned SMC1 ChIA-PET [42](**Figure S9A,C, Additional File1**). There was no significant correlation between ChAs values obtained for the ChIA-PET dataset and the promoter capture datasets (**Figure 2B**) showing that the ChAs measures are specific to the types of contacts assayed (**Figure S10-11, Additional File 1**). The cohesin ChIA-PET network is not enriched for promoters - only 20% of the SMC1 ChIA-PET fragments overlap the PCHi-C promoter fragments (**Table S1,**

**Additional File 1**), but it still shows the assortativity of PcG features and the actively elongating RNAPII-S2P.

To exclude the possibility that the correlation observed in the two promoter capture datasets was purely due to the experimental technique used to map the contacts, we also calculated ChAs for an RNAPII ChIA-PET dataset. Interactions involving RNAPII (8WG16 antibody, recognizing all variants) were detected in mESC [43] allowing us to analyse an RNAPII focused chromatin interaction network (**Figure S9B,D, Additional File1**). In addition, this network allowed us to further test the differences in ChAs of RNAPII variants, which we have observed to be reproduced in the PCHiC, HiCap and RNAPII ChIA-PET networks but not in the SMC1 ChIA-PET (**Figure S9-11, Additional File 1**). The RNAPII ChIA-PET network is obviously enriched in promoter interactions (58% of the RNAPII ChIA-PET fragments overlap PCHi-C promoter fragments, see **Table S1, Additional File 1**) but contrary to the PCHi-C and HiCap promoter-capture networks it contains only fragments in which RNAPII is bound. Similarly to what we found in the PCHi-C and HiCap networks, PcG proteins and associated histone marks show considerably high ChAs in the RNAPII ChIA-PET network, but lower than H3K4me1 (an enhancer specific mark) and the repressive mark H4K20me3 (**Figure S9B, Additional File 1**).

The ChAs of the non-specific RNAPII-8WG16 is quite low (0.07) in the RNAPII ChIA-PET network compared to all other features (mean 0.1) (**Figure S9B, Additional File 1**). A low ChAs is expected, given that fragments in this network are highly enriched for presence of this feature (84% of fragments have an RNAPII-8WG16 peak, abundance = 0.5). This leads to uniform levels of RNAPII abundance on the nodes and hence we do not observe any localization of the mark in specific areas of the contact network. Interestingly, we do observe higher ChAs for the elongating variant RNAPII-S2P (0.19 vs 0.07 for the RNAPII-8WG16) accompanied by a comparatively lower abundance (0.25 vs 0.5 for RNAPII-8WG16), suggesting that regions of the genome in which elongation takes place interact preferentially (**Figure S9B, Additional File 1**).

Overall, we observe a significant correlation of the RNAPII ChIA-PET ChAs values with PCHi-C (Pearson's R = 0.37, p-value = 1.01 x 10$^{-3}$, **Figure 2B and Figure S10C, Additional File1**) and an even better correlation with HiCap (Pearson's R = 0.59, p-value = 9.77 x 10$^{-9}$, **Figure 2B and Figure S11B, Additional File 1**), despite the drastically different topology **(Figure S11D, Additional File 1).**

Comparing the results of our approach using these four different networks we conclude that the methodology is able to identify the putative roles of specific epigenomic features in mediating different types of chromatin contacts. The high ChAs values of PcG and RNAPII are confirmed in different datasets but different features acquire different levels of ChAs and, potentially, different relevance in the different contact maps. Although PCHi-C, HiCap and RNAPII ChIA-PET are all enriching for interactions involving promoters, there are clear differences in the resulting networks. Notwithstanding the strong differences in topology and networks statistics between promoter-capture and ChIA-PET networks (**Figure S9 C-E, Additional File1**), we find higher similarity between the three promoter-enriched datasets (PCHi-C, HiCap and RNAPII ChIA-PET, **Figure S10-11, Additional File1**). The correlation between ChAs of promoter-capture networks is improved when looking at PCHiC and HiCap subnetworks that only include P-P contacts or P-O contacts (**Figure 2B and Figure S12, Additional File 1**). We therefore proceed with our goal to use ChAs to analyse the difference between interactions involving two promoters and interactions between promoters and other genomic elements.

## Distinct chromatin assortativity properties of contacts amongst promoters and between promoters and other elements

As mentioned above, the experimental design of promoter capture HiC (PCHi-C or HiCap) produces chromatin fragments of two kinds: Promoter (P) fragments are the ones that are captured in the experiment because they match a library of promoters and are therefore identified as baits; Other-end (O) fragments are chromatin fragments found to interact with the promoter baits.

We first investigated the differences in chromatin features associated to PCHi-C contacts involving two promoters (P-P) and contacts involving a promoter and an Other-end fragment element (P-O). We calculated feature abundance and ChAs values for two subnetworks: the P-P network and the P-O network (**Figure 3A, Figure S12, Additional File 1**). We combine these data in a comparative ChAs plot to directly assess the relationship between the ChAs of chromatin features measured in the two different subnetworks in PCHi-C (**Figure 3B**).

Strikingly, we find a number of features with very different values of ChAs in these two subnetworks. For example, in **Figure 3B** we see a group of features with positive ChAs in the P-P interactions, implying that these epigenomic features are found in promoters that contact each other, and negative ChAs in the P-O interaction network, implying that these features are usually not present on the other-end fragments that contact promoters. The features that have discordant signs of ChAs in the two subnetworks include many promoter-specific histone modifications and chromatin factors, specifically H3K4me3 (typically denoting active promoters), HCFC1 (transcription activator complex), SIN3A (transcriptional repressor complex), KDM2A (H3K26 demethylase), NMYC, OGT (histone acetyl transferase complex), H3K4me2 and H3K9ac (denoting active promoters)[38]. Features that have slightly higher or equal ChAs in the P-O interactions include CBX3 (the HPɣ implicated in elongation [44, 45]) and RNAPII-S2P. PCHi-C can only detect interactions involving at least one promoter. At the same time, most of the epigenetic features considered here are characteristic of promoters, due to the large bias in datasets available in the literature. Therefore, we are unlikely to find features with higher ChAs in P-O vs P-P contacts, which would lie at the upper left corner above the diagonal in **Figure 3B**. However, the features closer to the diagonal are features that are present in both P-P and P-O contacts. The PcG proteins and their associated histone marks are found very close to the diagonal on the comparative ChAs plot of **Figure 3B**, suggesting that they are found at both P-P and P-O contacts, together with H3K36me3 and the cytosine modifications 5hmC and 5fC.

The comparative ChAs plots for the HiCap datasets are very consistent with the PCHi-C ones (**Figure 3C, Figure S12, Additional File 1**) as shown clearly in a scatter plot of the difference of ChAs between P-O and P-P subnetworks in the two datasets (**Figure 3C**, further comparisons of P-P and P-O ChAs are shown in **Figure S12**). Interestingly, we observe substantially different ChAs scores for different RNA polymerase variants exclusively in P-O contacts, with elongating RNAPII having a ChAs 23 fold higher than the non-elongating forms (ChAs of RNAPII-S2P = 0.23 vs 0.01 for RNAPII-8WG16 (**Figure 3B**).

In order to assess the robustness of these differences, we generated 100 networks by random partial rewiring of the original network and re-calculated the ChAs in P-P and P-O subnetworks (see Methods and **Figure S12H)**. The simulations show non-overlapping simulated ChAs distributions in the P-O subnetworks for the different RNAPII variants, whereas the corresponding distributions in the P-P subnetworks are highly overlapping. These results suggest a significant difference in the role of elongating polymerase between P-P and P-O contacts.

## Characterization of overlapping chromatin communities reveals PcG and RNAPII-S2P modules

A large portion of the PCHiC interactions form a large connected component (LCC), also called a 'giant component' [35]. There is a significant correlation of the ChAs values measured for the LCC and for the interactions in the rest of the network (Pearson's r=0.8, p=0, **Figure S13, Additional File 1**). However, we observe a higher ChAs for PcG features in the LCC (mean 2.8 fold increase, especially EZH2 having ChAs = 0.37 in LCC compared to ChAs = 0.14 in the rest of the network). Considering the LCC we then identify features that are most abundant in nodes with high betweenness centrality, defined as the number of shortest paths from all nodes to all others that pass through that node [46]. PcG features are enriched in nodes with high betweenness centrality, again suggesting PcG's role in holding the core of the interaction network together (**Figure S14A, Additional File 1**).

To investigate whether PcG features were also involved in mediating connections between different chromatin communities, or neighbourhoods [35], we analysed the LCC with the ModuLand algorithm, which identifies overlapping modules [47] (**Figure 4A** and **Text S3, Additional File 1**). Once overlapping communities were defined, we calculated the 'bridgeness' of each node, defined as the number of different chromatin communities (modules) that it belongs to [48]. **Figure 4B** shows that the features most abundant in the nodes with highest bridgeness are the ones typical of promoters (SIN3A, HCFC1 and H3K4me3) as well as transcription factors such as E2F1, N-MYC, C-MYC, KLF4. On the contrary, PcG features are not abundant in high bridgeness nodes, suggesting that nodes in which PcG is present do not tend to belong to multiple chromatin communities.

The relative values of bridgeness and betweenness centrality can be used to distinguish the so-called "*date*" and "*party*" hubs, defined as nodes that entertain multiple interactions respectively one at a time or simultaneously [49, 50] (**Text S4, Additional File 1**). Extending this concept and using the enrichment of features in top bridgeness and betweenness nodes, we can identify 'date features', found in nodes that belong to multiple communities at the same time, and 'party features', found in nodes involved mainly in one community at any one time (**Figure S14B, Additional File1**). Only the PcG features (and to a lesser extent KDM2B, TAF1 and H4K20me3) appear to have a definite "party" character, suggesting that they might mediate more stable interactions, due to their high abundance in nodes that are central in the network (high betweenness) but mostly belong to a single community (low bridgeness, **Figure S14B**, **Additional File 1**). Similarly to what was observed for values of ChAs in the P-O subnetwork (**Figure 3B**), we see a striking difference between the elongating RNAPII variant S2P and non-elongating RNAPII variants (**Figure 4B, Figure S14B, Additional File 1**). The non-elongating RNAPII variants show similarly high abundance in top bridgeness and top betweenness nodes, suggesting their presence in nodes that are central and shared between multiple modules. On the contrary, the elongating S2P variant is found in more peripheral nodes that specifically belong to a single module, as shown by equally low enrichment in top bridgeness and top betweenness nodes (**Figure S14B, Additional File 1**). To summarize, PcG features are found in highly connected and highly central nodes, but these nodes do not belong to distinct network communities. The elongating variant of RNAPII, contrary to other RNAPII variants, is found mostly in nodes that belong to a single community and they are more peripheral to the network (low betweenness centrality).

We investigate the difference between RNAPII variants further by looking at enrichment of features in chromatin communities identified by ModuLand, concentrating on the features that showed a high value of ChAs (ChAs > 0.1, **Figure 4B**). The heatmap in **Figure 4C** clearly shows the presence of 4 clusters. The largest and more prominent is cluster IV including all PcG features, which are enriched in a specific subset of chromatin communities. Clusters II and III contain respectively non-elongating forms of RNAPII and DNA cytosine modifications. On the other hand, RNAPII-S2P appears in cluster I in chromatin communities that are also enriched in H3K36me3 and CBX3. Although all enrichments in RNAPII are anti-correlated with enrichments in PcG features (**Figure 4C**), this anti-correlation pattern is stronger for the actively elongating variant RNAPII-S2P (**Figure S15, Additional File 1**). Overall, these results suggest that PcG features are found in very central and connected nodes that interact stably forming specific chromatin communities. Similarly, active elongation is taking place in specific chromatin communities but fragments of chromatin bound by elongating RNAPII are not particularly connected or central in the network **(See also Figure S6, Additional File 1)**. In the next section we will explore the differences between the different RNAPII variants in more detail.

## RNAPII-S2P has higher ChAs in promoter-other end contacts compared to other RNAPII variants

Our collection of genome-wide features includes five different ChIP-seq datasets for RNAPII obtained using different antibodies. Of these, three of them recognize different phosphorylated forms of RNAPII involved in the different stages of transcription [51, 52] (**Figure 5A**). We can therefore distinguish between ChIP-seq peaks of RNAPII in its initiating or repressed form (S5P, S7P), in its actively elongating variant (S2P), or in any of its variants (RNAPII-8WG16, POLII).

We compared the ChAs of the different RNAPII variants in the whole PCHi-C and HiCap networks. As was already noted, RNAPII-S2P, which denotes elongation of actively transcribed genes, shows higher ChAs than the other RNAPII variants in both datasets (**Figure 5B**). These differences are robust to partial rewiring of the networks (see Methods and **Figure S16A)**. **Figure 5C** shows the corresponding abundance values, which are very comparable between different RNAPII variants within each dataset.

Next, we compared the ChAs of the different RNAPII variants in the RNAPII ChIA-PET network (**Figure 5B**). In principle, the RNAPII ChIA-PET dataset provides us with the network of chromatin contacts in mESC mediated by any RNAPII, as the antibody used in this experiment (8WG16) recognizes all RNAPII variants. Interestingly, there is an increase of ChAs from repressed to actively elongating RNAPII in all three networks (**Figure 5B and Figure S16A**). These results suggest that, whereas all interacting fragments in these promoter-rich networks do contain some form of polymerase, the presence of active forms of RNAPII distinguishes different network neighbourhoods in which active elongation is taking place, as also suggested in **Figure 4C**.

Finally, we used the ChIA-PET network of contacts mediated by cohesin in mESC cells as a negative control [42]. In this dataset we see many contacts that do not involve any promoters or genes, in which we do not expect to find any RNAPII bound (61% of fragments in the SMC1 ChIA-PET dataset have no signal for RNAPII-8WG16). Indeed, the different variants of RNAPII in this cohesin-mediated network have very high ChAs (**Figure 5B and Figure S16A**). The presence of any form of RNAPII clearly separates regions of the cohesin-centered network where transcription is active from regions where it is not. These trends cannot be explained by changes in abundance (**Figure 5C**).

We further compared the ChAs of different RNAPII variants between P-P and P-O contacts (**Figure 5D**). In the PCHi-C network we observe the ChAs for different phosphorylation states of RNAPII to vary widely in the P-O contacts (from close to 0.01 to 0.23, the third highest value overall) while all states have similar ChAs in the P-P contacts ( ChAs range 0.21-0.22, **Figure 5D and Figure S16B, Additional File1**). To understand this trend better, we also look at abundance of the different RNAPII variants in the different subnetworks (**Figure S16C, Additional File 1**). Whereas in the P-P subnetwork the abundance decreases from inactive forms of RNAPII to the elongating form, in the P-O subnetwork the elongating form is equally abundant compared to the other forms. We can therefore conclude that the different ChAs observed for different forms of RNAPII is related to the topological distribution of RNAPII binding on the network, rather than simply to changes in average abundance in the network. This finding suggests that when O fragments contact P fragments, predominantly the elongating form of RNAPII is present on both fragments. The difference between different RNAPII forms specific to P-O contacts is even more evident in the HiCap dataset where the ChAs value of non-elongating variants of RNAPII is negative (**Figure 5E, Figure S16D, Additional File1**). This is likely due to the higher resolution of the HiCap experiment, which allows us to better discriminate P and O fragments that are probably merged in some of the larger PCHi-C fragments.

We investigated further to determine whether the patterns of ChAs of different RNAPII variants change depending on the type of fragments contacted by the promoter. We selected two types of O fragments: enhancers (fragments with H3K4me1>0) divided into active enhancers (H3K4me1>0 and H3K27ac>0) and poised enhancers (H3K4me1>0 and H3K27me3>0). We can thus separately compare ChAs values between P-P contacts and contacts of P fragments with each type of O fragment. As shown in **Figure 5F**, RNAPII-S2P has higher ChAs than the other RNAPII variants in contacts between promoters and active enhancers, but not in contacts with poised enhancers (**Figure S16, Additional File 1**). This suggests that the presence of elongating RNAPII at the P-O contact and the activity of the enhancer might be related.

Strikingly, we also observe a considerable number of contacts between promoters and fragments that do not have the H3K4me1 enhancer mark (H3K4me1=0, referred to as non-enhancers in the figure), which we found to be strongly enriched for H3K36me3 and that in 19% of cases overlap protein coding gene bodies (**Figure S17, Additional File 1**). In these contacts ChAs varies from very negative in the non-specific forms to highly positive for the elongating form. This is not due a change in the abundance of different forms of RNAPII (**Figure S18, Additional File 1**) and these results are largely confirmed in HiCap (**Figure 5G**). These findings suggest that promoters can contact transcribed gene bodies.

## Discussion

**Assortativity as a robust approach to identify important features in chromatin contacts**

We have presented a novel approach, inspired by social network science, which enables the powerful integration of epigenomic features with maps of 3D contacts of chromatin fragments in the nucleus, taking into account the exact network topology. Our approach is robust to the random removal of edges in the contact map, due to its global character.

Using the PCHi-C network in mESCs we demonstrated the capabilities of our assortativity based approach in recapitulating the importance of PcG factors and associated histone marks. Given the small proportion of fragments that are covered by these marks in the whole genome, the values we observe for their ChAs are highly significant, as also shown by two different randomization procedures. Most features show no change in ChAs value when considering only long-range interactions. PcG features even show higher assortativity in the long-range subnetwork, which is consistent with recent results about PcG mediating extremely long range contacts [20].

So far, integrated analyses report correlations between genomic information and characteristics of genes in the 3D contact network [4, 10, 53–55], but the exact network topology itself is rarely taken into account. On the contrary, the network topology is part of the definition of ChAs and has direct implications in the subsequent calculations. Two very inspiting recent works predict 3D interactions based on 1D epigenomic profiles, but neither provides major insight on the network topology [12, 56].

Having ascertained the appropriateness of chromatin assortativity as a measure, we further define two different subnetworks formed by P-P and P-O interactions and then compare the ChAs for all the features in the two subnetworks. These comparisons show the specific association of certain chromatin features with each type of contact. For example, H3K4me3 has a low ChAs in the complete network but high ChAs in the P-P subnetwork and negative ChAs in the P-O contacts, as corresponds with its role as a differential mark of active promoters.

The *ChAs difference* between the two types of contacts summarises the relationship between features and network topology and permits a direct comparison between datasets. For example the comparison of ChAs scores between the promoter-capture and the ChIA-PET datasets shows how our method can identify very specific characteristics of the chromatin interaction networks and expose experimental biases. Furthermore, it could be used to identify low quality ChIP-seq datasets, which would fail to show the expected ChAs values.

**Biological interpretation of ChAs**

We performed this comparison using PCHi-C and HiCap networks to exclude the possibility that our findings are artefacts of one specific dataset. We find a strong correlation of the ChAs values between P-O and P-P subnetworks in the two datasets, giving us confidence in the biological relevance of our results. The reproducibility between the two datasets is remarkable, especially considered the differences in the experimental techniques and the interaction filtering methods used. Whereas PCHi-C is enriched for long-range contacts, HiCap has a higher coverage of short-distance interactions [5, 7, 26], likely constituting connections between promoters and regulatory elements that are relatively close. These types of interactions are probably lost in PCHi-C, due to the larger fragment size (which means a single fragment might contain both sites of interaction), and due to the strict distance correction algorithms applied [29]. Given these differences, the good correspondence of ChAs in the two datasets suggests a general importance for many chromatin factors, which seem to play similar roles in short and long-range contacts. This is consistent with our observation that ChAs of most features is maintained when removing short-distance contacts (**Figure S5, Additional File 1).** There are however very interesting differences between the ChAs values in P-O contacts in PCHi-C and HiCap, which can be seen by comparing ChAs values directly in the two datasets. For example, more features have negative P-O ChAs values in HiCap. The reason for this is that in PCHi-C the larger fragments will include promoters and also nearby regulatory regions, decreasing the difference between P and O fragment associated chromatin features.

Looking at the P-P and P-O subnetworks separately also allowed us to notice a marked difference between the variants of RNAPII. The elongating variant appears more strongly associated with contacts between promoters and active enhancers or transcribed gene bodies, compared to inactive forms. This is observed in all promoter centred interaction datasets, including PCHi-C, HiCap and RNAPII ChIA-PET. In fact this tendency is given by a decrease in assortativity of the non-elongating RNAPII forms in the contacts between promoters and active enhancers or transcribed gene bodies.

Recently, the presence of RNAPII at distal sites was functionally linked to the activity of CEBP-bound enhancers, showing that active binding sites display stronger RNAPII binding and local enhancer RNA production [57]. The presence of polymerase at enhancers was also shown to be strongly predictive for the timing of enhancer activation during development [58]. Our analysis goes beyond these findings and suggests that the presence of non-elongating variants of RNAPII is not associated to preferential contacts of promoters and active distal regulatory elements, whereas the elongating form is. This picture is also consistent with the negative ChAs of non-elongating forms of RNAPII in HiCap P-O contacts. It is possible that the RNAPII that is found at active enhancers is mostly in its elongating form. This is also confirmed by looking at the abundance of RNAPII variants in different fragment types (**Figure S18, Additional File 1**) which shows that the only form of RNAPII present on other elements is the elongating one. The result is stronger in HiCap contacts, probably because the large size of PCHi-C fragments might signal peaks of RNAPII in O fragments where in reality the peak is in a nearby promoter.

These results are consistent with the different distribution observed between the elongating and non-elongating forms of RNAPII across chromatin communities. Many nodes of the network are found to belong to multiple communities, as evidenced by their high bridgeness. This could indicate that these fragments tend to interact with different partners, either in time or in different cells of the population assayed [31]. The low bridgeness of the elongating form suggests that fragments that are being actively elongated mostly stay within a single chromatin community. Moreover, these fragments are likely to be peripheral to the community itself, given the low betweenness of nodes with high abundance of RNAPII-S2P.

This interpretation would be in agreement with the stationary model for RNAPII in transcription factories, (assemblies of genes being co-transcribed [16, 59–61] where elongating RNAPII and nascent transcripts would be localized at the periphery of factories.

We estimated the PcG features to have a more 'party-hub' character rather than 'date-hub', given the abundance of these features in top betweenness and top bridgeness nodes [46, 48]. The concept of date and party hubs is better defined for dynamical networks, typically protein interaction networks in which the former type refers to one-to-one interactions and the latter to stable complexes [48, 62]. In our case we can speculate on the meaning of this distinction, suggesting that PcG features are associated to more stable contacts, which could be more stable both in time and across different cells in a large population [17], and span longer chromosomal distances [20]. On the contrary, features present in active promoters and mediating promoter-enhancer contacts are likely to be more specific. The peculiar characteristics of contacts mediated by PcG could be related to the recent observation of major differences between chromatin in the PcG repressed or poised state [63]. These super-resolution microscopy studies found the PcG chromatin to be differently packed from fully active or repressed chromatin, suggesting that the poised domains spatially exclude neighbouring active regions.

To summarise these results, we propose the model in **Figure 6**, where the network of chromatin contacts (sketched in **Figure 6A**) shows regions of promoters that are active, probably due to their contacts with active regulatory elements and transcribed gene bodies. This would lead to high ChAs for the elongating form of RNAPII in both P-P and P-O contacts while ChAs of non-elongating forms would stay low in P-O contacts. Recent literature is suggesting a picture in which enhancer activity is mediated by the formation of loops connecting the gene promoter, the distal enhancer and the body of the gene [15, 16]. Moreover, 3C experiments have shown that these gene-body contacts are often dynamic and they keep a connection between the gene promoter and the gene body at the exact location of active elongation [64]. We suggest that the RNAPII-S2P variant might be involved in these contacts (**Figure 6B**). In the fruit fly, it was proposed that promoter-enhancer contacts are preformed, conserved across tissues and developmental stages, and associated with paused RNAPII[65]. Further experiments will be needed to assess the role of elongation in these processes.

The many interactions we have observed between promoters and gene body fragments without the H3K4me1 enhancer mark cannot be easily explained. It could be speculated that these contacts are joining two promoters while both genes are being transcribed, such that each promoter could come in contact with the body of any of the two genes. This scenario would be consistent with the concomitant localized transcription of multiple genes. This picture is again in line with the concept of transcription factories. Our results on RNAPII-S2P further corroborate this model and are consistent with experimental results showing that whereas the RNAPII-S5P variant would accumulate in the factory, the RNAPII-S2P would remain in the nuclear space nearby the factory [66]. The co-enrichment of modules that we observe for RNAPII-S2P, H3K36me3 and CBX3 (which was shown to interact physically with CDK12 [67], which in turn produces the phosphorylation on RNAPII necessary for active elongation) further support the separation of fragments being actively elongated from the transcription factory.

## Conclusions

We have demonstrated the use of assortativity of chromatin features in interpreting chromatin interaction datasets in the context of available epigenomic data. We have achieved this through the definition of ChAs, a measure of how much the value of a specific chromatin feature is correlated between a chromatin fragment and others that interact preferentially with it. The difference of ChAs between the P-P and P-O subnetworks can be used to compare two or more chromatin interaction datasets. Thus the method we present provides a quick and efficient comparison of different chromatin interaction networks and integration of complementary epigenomic and functional information.

Comparing two different networks obtained with two variations of promoter capture HiC on mESC cells, we find excellent reproducibility of the following observations: 1) Members of the PcG and associated marks show high ChAs despite the low abundance in the interacting fragments, suggesting that they mediate the 3D contacts, especially in the long range, as already noted [20]; 2) The ChAs values of different variants of RNAPII suggest a picture in which contacts happen between enhancers, their target promoters, and along the gene body. Moreover, we identify the important role of the actively elongating variant of RNAPII in interactions between promoters, distal elements and other sites in the gene body. Whether it is the contact between these different chromatin regions that spreads the localization of RNAPII-S2P or RNAPII in its elongating form that promotes the contacts remains to be examined in further work. ChAs is a new complementary measure that provides a global view based on integrating network topology with feature values in the interacting fragments. It has recently been suggested that features located within the loop connecting promoter and enhancer can be determining for the interaction [12], which suggests that expanding our analysis to combine HiC and Promoter Capture HiC derived networks might yield further insight.

Our results across four different chromatin interaction networks, spanning different techniques and identifying different biases, lend support to the presented ChAs approach as a useful tool in the quest for organizing principles shaping chromatin contact networks.

## Methods

### Generating the PCHi-C network

PCHiC Interactions measured in mESCs in [8] were processed using CHiCAGO [29]. The publicly available HiCUP pipeline (Wingett et al. Submitted) was used to process the raw sequencing reads. This pipeline was used to map the read pairs against the mouse (mm9) genome, to filter experimental artefacts (such as circularized reads and re-ligations), and to remove duplicate reads. The resulting BAM files were processed into CHiCAGO input files, retaining only those read pairs that mapped, at least on one end, to a captured bait.

CHiCAGO is a method to detect significant HiC interactions specifically adapted to promoter capture experiments. In brief, it uses a noise convolution model in which two noise terms account independently for noise sources that dominate at different scale: 1) Brownian motion which leads to probabilities of interactions decreasing with distance and is modelled using a negative binomial distribution and 2) sequence artefacts which are modelled using a Poisson distribution. Once the ChiCAGO scores had been obtained only interactions with score>= 5 were considered.

The network was then built considering each fragment as a node (therefore having two types of nodes, namely promoters and other ends, and two types of edges, namely promoter-other end and promoter-promoter. Multiple edges connecting the same 2 nodes were eliminated.

**HiCap and ChIA-PET networks:**

The HiCap data was downloaded from the supplementary material of Sahlén et al. [7] which provides coordinates of the promoter and other end fragments that show significant interaction as well as a list of gene promoters that interact based on assignation of promoter fragments to the closest TSS. Interactions not involving promoter fragments were filtered out. The fragments coordinates and interactions of the SMC1 ChiA-PET dataset were downloaded from the supplementary material of [42]. The fragment coordinates of the RNAPII ChIA-PET dataset were downloaded from the supplementary material of [43]. No further processing or filtering was made for these two datasets.

**Calculation of feature abundance in the chromatin fragments:**

The chromatin features (Additional File 2) were taken from Juan et al. [38] and the binarization was performed as described there in 200bp windows. For each fragment the overlapping windows of chromatin peaks were identified and their values averaged to give a percentage of presence of any feature in each fragment. Thus for each feature a value between 0 and 1 is associated to each fragment (which has an average length of 4.9Kb in PCHiC and 600bp in HiCap) generating a fragment-by-feature matrix. The value of abundance of a feature is defined as the average of that feature value across all fragments in the network considered. The feature abundance for each fragment is listed in Additional File 6.

**ChAs calculation:**

We define the ChAs of a specific epigenomic feature in a contact network as the Pearson correlation coefficient of the value of that feature across all pairs of nodes that are connected with each other [40]. ChAs is therefore the assortativity of the abundance value of a feature on a network. We used the igraph (Version 0.7.1) package in R and its function 'assortativity' to calculate the ChAs of each feature separately on the network of choice (either the full network, the P-P or the P-O networks in PCHi-C and HiCap.

The assortativity measure used was that for continuous variables given by the following formula:

$$r = \sum_{jk} jk(e_{jk} - q_j q_k)/\sigma_q^2$$

Where $q_i = \sum_j e_{ij}$, $e_{ij}$ is the fraction of edges connecting vertices of type $i$ and $j$, with $\sigma_q$ is the standard deviation of $q$.

A more intuitive definition of assortativity is simply the Pearson correlation between two vectors: vector 1 contains the feature values of the source nodes and vector 2 contains the feature values of the sink nodes, once all edges in the network are enumerated. There is no appreciable difference in the value of assortativity obtained by listing all edges in an arbitrary direction or first adding all edges in the opposite direction and calculating assortativity on this extended network.

### Robustness and significance of ChAs values:

We assessed how the ChAs values can be affected by the accuracy of the topology of the chromatin interaction network by removing edges at random and following targeted approaches based on node degree, node centrality and feature abundance. Further details and results can be found in **Text S1** and **Figure S2, Additional File 1**.

We also tested whether the ChAs of the chromatin features we measured was significantly higher than would be expected at random using two different approaches. Briefly, in the first approach we shuffled the assignment of feature values between network nodes, repeating this 100 times and thus calculating empirical p-values. In the second approach we created new interactions between bait fragments of Chromosome 1 and randomly chosen regions of the same chromosome, with the same size and distance from bait as the original other-end fragments, also calculating empirical p-values. Further details, a schematic description of the two approaches and results can be found in **Text S1** and **Figures S3-4, Additional File1**.

Finally, to assess the impact of differences between ChAs of different features in the same network or the same feature across networks, we performed a partial rewiring of the networks and calculated the distribution of ChAs values for each feature (10% edges swapped.

### Network analyses and community detection

Network properties such as degree, transitivity, betweenness centrality and number of connected components were calculated using igraph. Further details on the network analyses and results can be found in **Text S2** and **Figure S6, Additional File 1**.

We identified chromatin communities in the PCHi-C network using two separate approaches. First, we used the ModuLand plugin for Cytoscape [47], which returns overlapping network communities and values of bridgeness for each network node (defined as number of communities that the node belongs to [46]). Second we used a fast greedy community finding algorithm from the igraph package to identify non-overlapping network modules. Further details on the community detection and results can be found in **Text S3** and **Figure S7, Additional File 1**.

### Definition of different types of O fragments

Active enhancers were defined as other-end fragments with the H3K4me1>0 and H3K27ac>0. Poised enhancers were defined as other-end fragments with the abundance of H3K4me1>0 and H3K27me3>0. For example, given our definition of feature abundance, this will identify an active enhancer in any fragment that has at least one 200bp segment covered by a H3K4me1 peak and at least one (not necessarily the same) 200bp segment covered by a H3K27ac peak. We have identified Non-enhancers as O fragments that have H3K4me1=0.

### Statistical analyses

All analyses were performed using R version 3.1.0 (x86_64-pc-linux-gnu)(R Development Core Team 2008).

## Availability of supporting data

The chromatin features used were already compiled from the literature in [38] and can be found in Additional File 2. A table with calculated values of abundance and ChAs for all the networks used and all the features can be found in **Additional File 3.** Results of the randomization approaches can be found in **Additional Files 4-5** and abundances for fragments in the different networks can be found in **Additional File 6.** Additional files, the code and networks are available at https://figshare.com/s/c331ded300fefd8f3ec0.

## List of abbreviations:

**ChIA-PET** Chromatin Interaction Analysis by Paired-End Tag sequencing

**PCHi-C** Promoter Capture HiC

**RNAPII** RNA Polymerase II

**mESC** mouse embryonic stem cells

**PcG** Polycomb Group

**ChICAGO** Capture Hi-C Analysis of Genome Organization

**P** Promoter

**O** Other-end

**P-P** Promoter-Promoter

**P-O** Promoter-Other end

**LCC** Largest Connected Component

**ChAs** Chromatin Assortativity

# Competing Interests

The authors declare that they have no competing interests.

# Authors' contributions

VP performed the analyses; EC, VP and MS processed the different datasets; VP, EC, BMJ, DJ and DR interpreted the results. VP and DR devised the study and drafted the initial manuscript. DR, MS, PF and AV supervised the work. All authors revised and approved the final version of the manuscript.

# Description of additional data files

The following additional files are available with the online version of this paper:

Additional File 1 is a PDF file containing all supplementary texts, figures and tables.

Additional File 2 is an .xls file containing all feature used and references to the datasets [38].

Additional File 3 is a tabbed formatted text file containing the calculated values of abundance and ChAs for all features in all 4 networks.

Additional File 4 is a tabbed formatted text file containing the results of the PCHi-C randomization preserving network topology.

Additional File 5 is a tabbed formatted text file containing the results of the PCHi-C randomization preserving feature distributions along the chromosomes.

Additional File 6 in an Rdata file containing all feature values on each fragment in the 4 networks.

# Acknowledgements

The research leading to these results has received funding from the European Union's Seventh Framework Programme (FP7/2007-2013) under grant agreement number 282510 (BLUEPRINT) and in the framework of the Platform for Biomolecular and Bioinformatics Resource PT 13/0001/0030 of the ISCIII, which is funded through the European Regional Development Fund (ERDF). This work was also supported by grants from the Biotechnology and Biological Sciences Research Council BBS/E/B/000C0404 and the Medical Research Council MR/L007150/1, to PF. VP acknowledges a FEBS Long-term fellowship. We also thank Simone Marsili and two anonymous referees for providing interesting comments and improvements to our work.

# Figure Legends

**Figure 1. Illustration of chromatin assortativity (ChAs) of epigenomic features in a network of chromatin contacts. A)** PCHi-C chromatin interaction network in mESCs. Nodes are coloured by proportion of fragment covered by EZH2, which highlights the neighbourhoods in which the protein is abundant. B) Visualization of the genomic region highlighted in the box (A) using the WashU Epigenome browser [68] with added custom tracks for PCHiC interactions and EZH2 peaks together with othe PcG related features. C) Cartoon illustrating what ChAs measures. Each node of the network is a chromatin fragment, blue nodes represent nodes in which a peak of a specific chromatin mark is found and edges represent significant 3D interactions. Next to it we show a cartoon plot of ChAs versus abundance.

**Figure 2. ChAs in the PCHi-C network and correlations of ChAs values in other networks.** A) ChAs of the 78 chromatin features in the PCHi-C Chromatin Interaction Network. For clarity, some feature names have been omitted; see Additional File 3 for the correspondence between features and numbers. B) Comparison of the correlations of the ChAs values yielded by PCHi-C, HiCap, RNAPII ChIA-PET (ChIA.RNAPII) and SMC1 ChIA-PET (ChIA.SMC1). Ellipse width and color is proportional to the Pearson's R coefficient (see color legend). Only p-values > 0.01 are shown.

**Figure 3. Comparing the assortativity of Promoter-Promoter (P-P) and Promoter-Other end (P-O) contacts.** A) Schema of a full network (left) that is de-composed into a P-P network and a P-O network (right). B) PCHi-C Compartive ChAs in P-O versus P-P subnetwork. C) Differential subnetwork ChAs in HiCap versus PCHi-C.

**Figure 4. Chromatin communities in the largest connected component (LCC) of the PCHi-C network.** A) Chromatin communities are defined based on the connectivity using ModuLand, which outputs overlapping communities[47]. The bridgeness of a node indicates the number of communities it belongs to. B) ChAs in LCC vs enrichment in in the top 500 bridgeness nodes for all features, PcG and RNAPII features are shown with bigger circles. C) Hierarchical clustering of the empirical p-value of the enrichment for the top ChAs features (ChAs > 0.1) along the chromatin communities.

**Figure 5. ChAs of different RNAPII variants in promoter-capture and ChIA-PET networks.** A) Different variants of RNAPII in our chromatin feature set. B) Comparison of ChAs of RNAPII in PCHi-C, HiCap, RNAPII and SMC1 ChIA-PET subnetworks. C) Comparison of abundance of RNAPII in PCHi-C, HiCap, RNAPII and SMC1 ChIA-PET subnetworks. D) PCHi-C ChAs in P-P and P-O subnetworks. E) HiCap ChAs in P-P and P-O subnetworks. F) PCHi-C ChAs compared in P-P and different types of P-O subnetworks. G) HiCap ChAs compared in P-P and different types of P-O subnetworks.

**Figure 6. Schematic representation of a model of chromatin fragments interactions.** A) Interpretation of the ChAs results for RNAPII and RNAPII-S2P with a cartoon network in which we highlight P-P and P-O contacts. The elongating variant RNAPII-S2P is associated with active enhancers which contact promoters that might also contact each other. This situation corresponds to equally high ChAs in both types of contacts. Other forms of RNAPII (lacking S2 phosphorylation) have lower ChAs in P-O contacts but high ChAs in P-P contacts. B) Virtual-4C (extraction of interactions centred on specific genomic location from genome-wide data) anchored on the HOXA1 promoter showing P-P and P-O contacts and corresponding peaks of different RNAPII variants. C) A model of loops formed between distal regulatory elements, promoters and gene bodies bound by RNAPII-S2P.

A

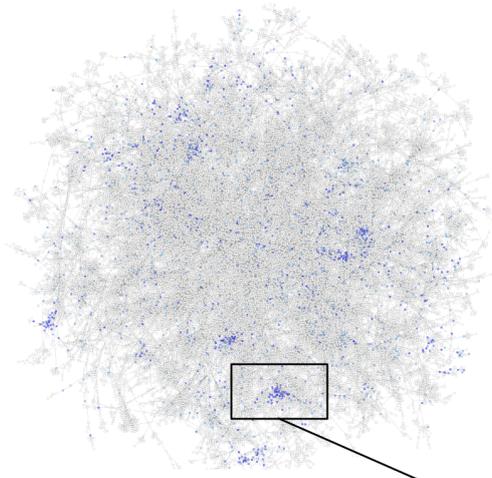

B

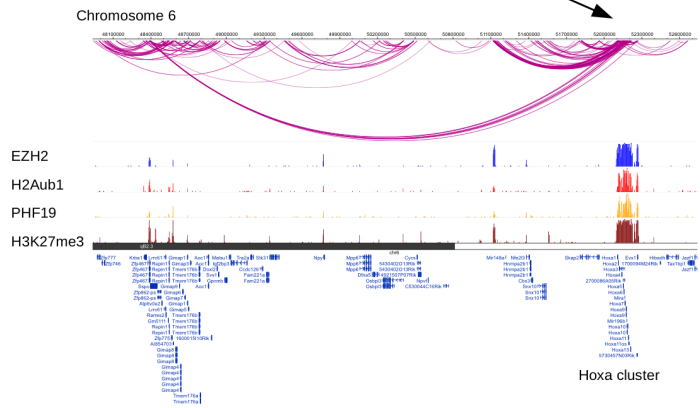

C

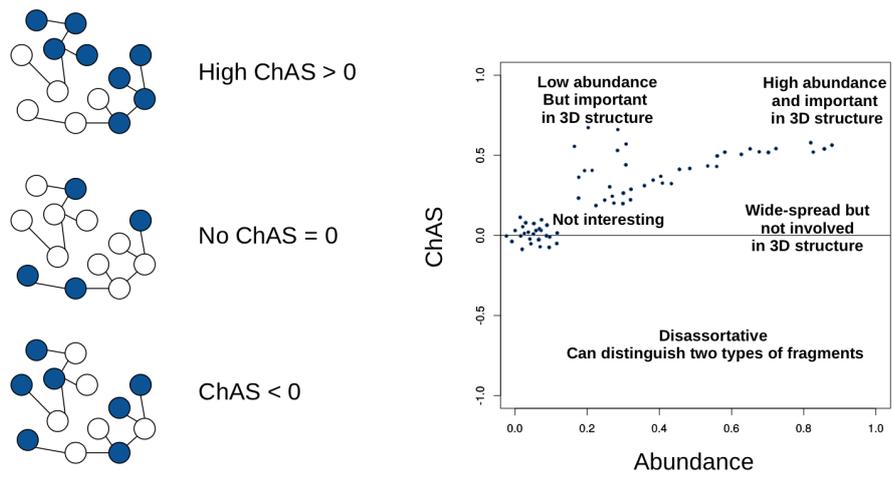

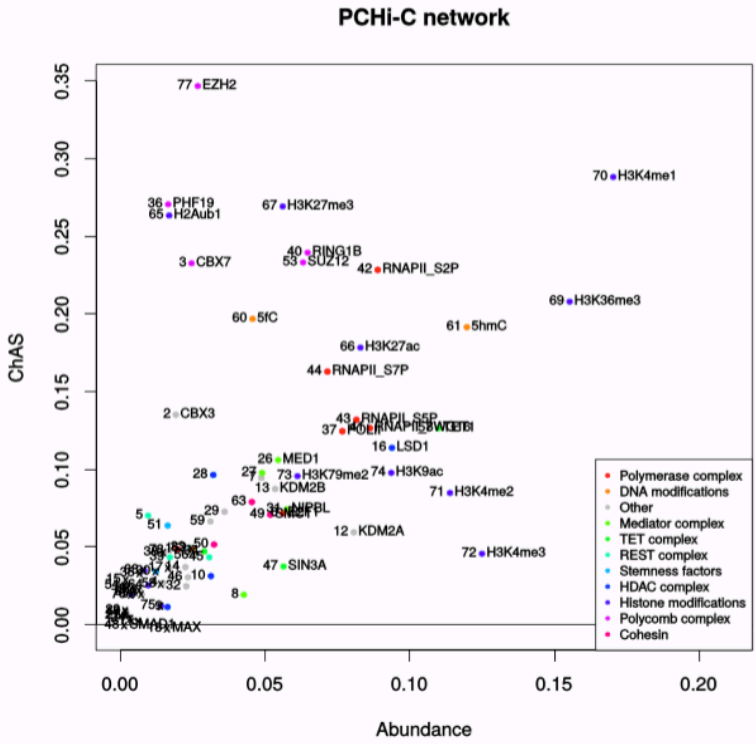
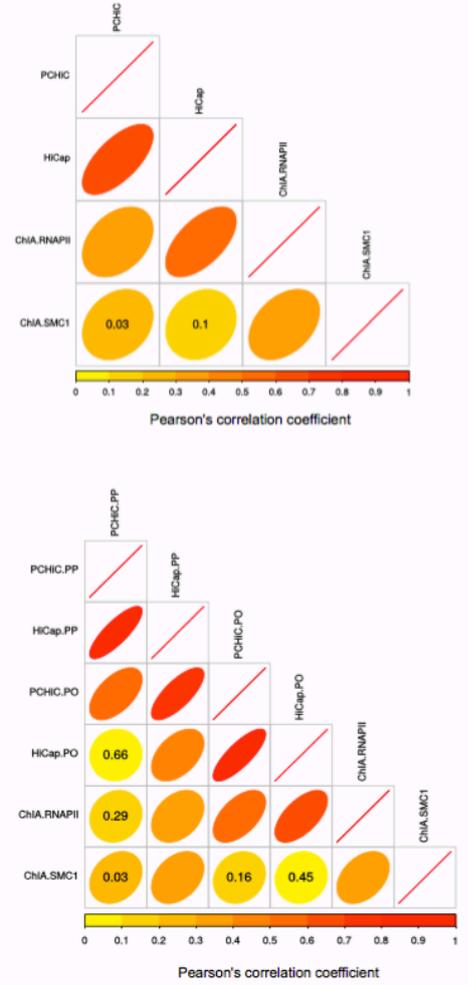

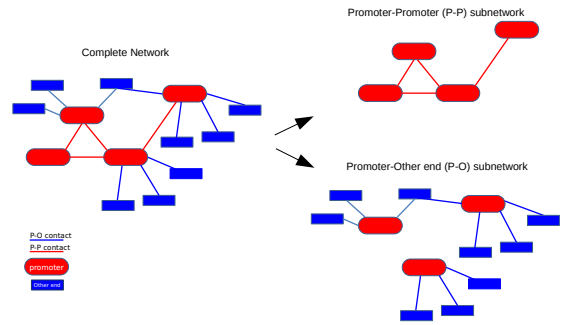

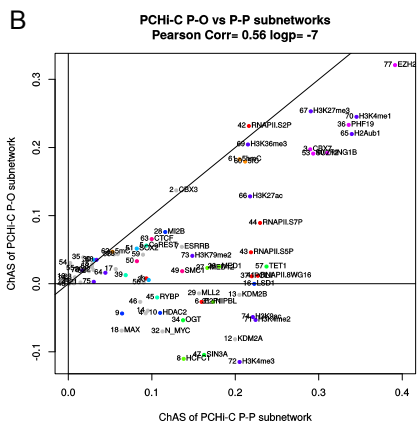 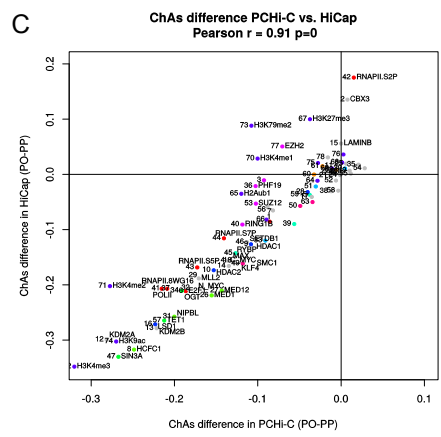

A

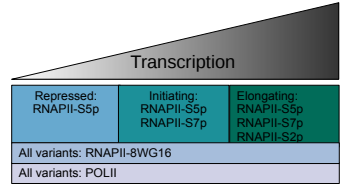

B 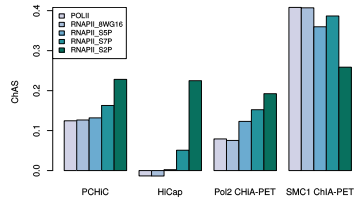 C 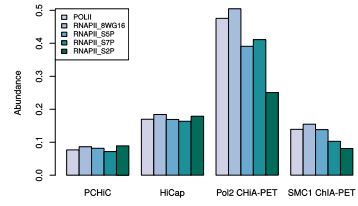

D 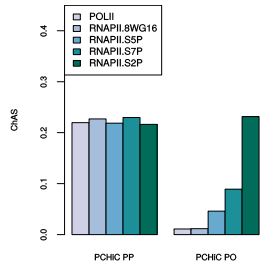 E 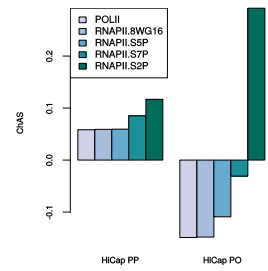

F 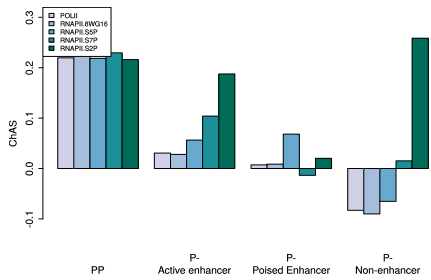 G 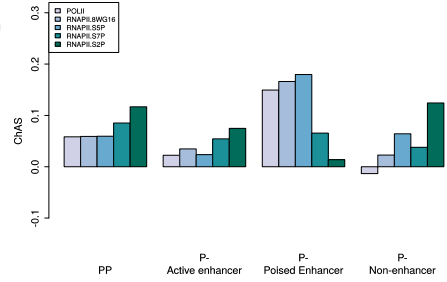

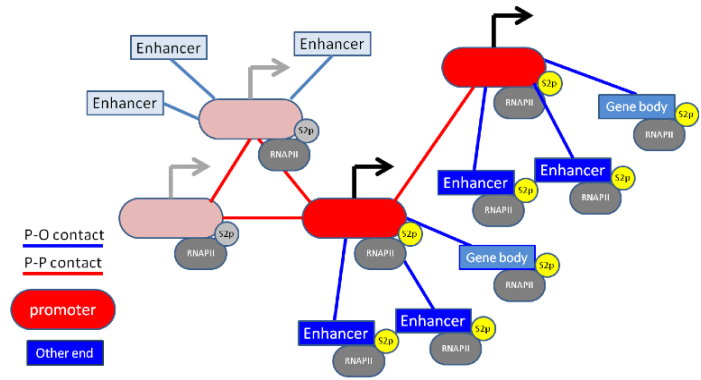

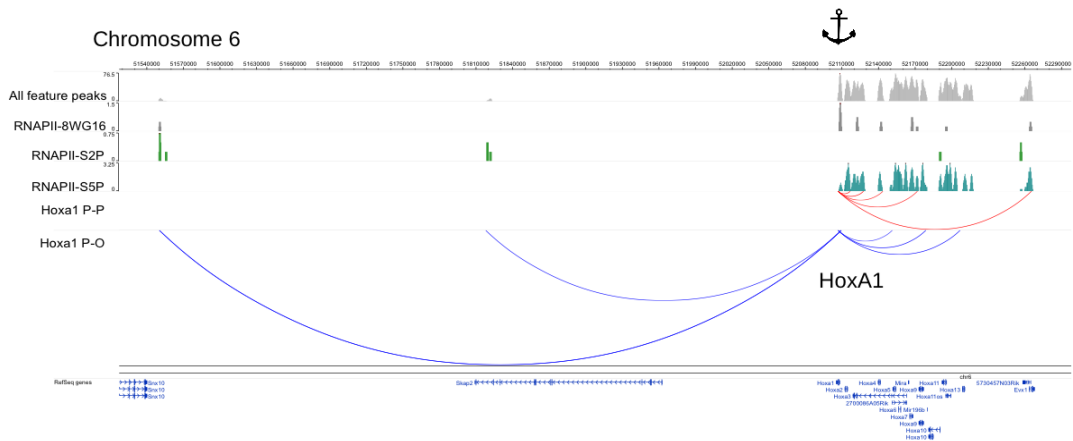

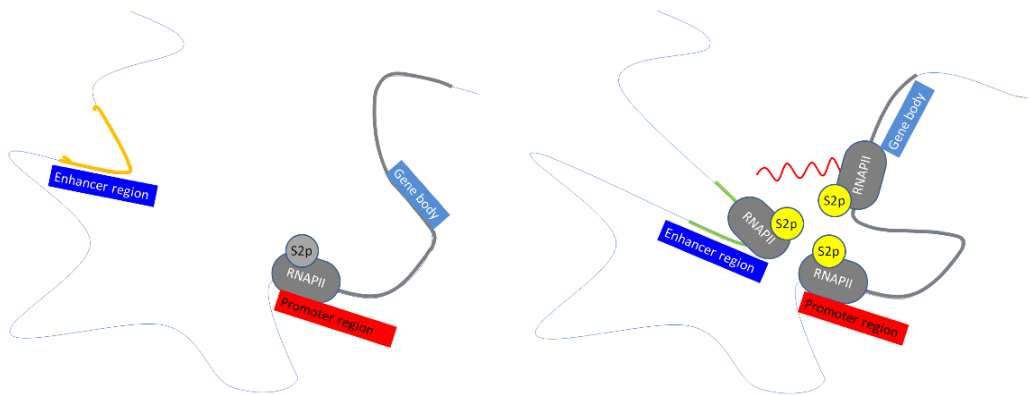